\documentclass[twocolumn,pra,showpacs,superscriptaddress]{revtex4}

\usepackage{graphicx}
\usepackage{dcolumn}
\usepackage{bm}

\begin{document}

\title{Josephson dynamics of a spin-orbit coupled Bose-Einstein condensate \\
 in a double well potential}

\author{Dan-Wei Zhang}
\affiliation{Laboratory of Quantum Information Technology and SPTE,
South China Normal University, Guangzhou, China}

\author{Li-Bin Fu}
\affiliation{Science and Technology Computation Physics Laboratory,
Institute of Applied Physics and Computational Mathematics, Beijing
100088, China}

\author{Z. D. Wang}
\affiliation{Department of Physics and Center of Theoretical and
Computational Physics, The University of Hong Kong, Pokfulam Road,
Hong Kong, China}

\author{Shi-Liang Zhu}
\email{slzhu@scnu.edu.cn} \affiliation{Laboratory of Quantum
Information Technology and SPTE, South China Normal University,
Guangzhou, China}

\affiliation{ Center for Quantum Information, IIIS, Tsinghua
University}

\begin{abstract}
 We investigate the quantum dynamics of an
experimentally realized spin-orbit coupled Bose-Einstein
condensate in a double well potential. The spin-orbit coupling can
significantly enhance the atomic inter-well tunneling. We  find
the coexistence of internal and external Josephson effects in the
system, which are moreover inherently coupled in a complicated
form even in the absence of interatomic interactions. Moreover, we
show that the spin-dependent tunneling between two wells can
induce a net atomic spin current referred as spin Josephson effects.
Such novel spin Josephson effects can be observable for
realistically experimental conditions.

\end{abstract}
 \maketitle

\section{introduction}
Based on the Berry phase effect \cite{Berry,Sun} and its
non-Abelian generalization \cite{Wilczek}, the creation of
synthetic gauge fields in neutral atoms by controlling atom-light
interaction has attracted great interest in recent theoretical
studies
\cite{GF_review,GF1,GF2,GF3,GF4,GF5,GF6,GF7,GF8,GF9,GF10,GF11,GF12,GF13,GF14},
and has  been realized in spinor Bose-Einstein condensates (BECs)
in the pioneering experiments of the NIST group
\cite{NIST_1,NIST_SOC} and also in several subsequent
experiments of other groups \cite{jingzhang,Bloch2011,Struck}. The
neutral atoms in the generated effective Abelian and non-Abelian
gauge fields behave like electrons in an electromagnetic field
\cite{NIST_1,jingzhang,Bloch2011} or electrons with spin-orbit
(SO) coupling \cite{NIST_SOC}. Different from electrons that are
fermions, the atoms with the synthetic SO coupling can be bosons and
typically BECs. This bosonic counterpart of the SO coupled
materials has no direct analog in solid-state systems and thus has
received increasing attention
\cite{Galitski,Wang,Jian,Ho,Congjun,Yip,Xu,Machida,Duine,Sarang,Santos,Hu,
Biao,DaSarma,SuYi,Han,Congjun_2,Galitski_2,Merkl,Zhang,Zhu} for
different types of SO coupling, different internal atomic
structures (pseudospin-1/2, spin-1 and spin-2 bosons, etc.), and
different external conditions (homogenous, trapped and rotated).
These theoretical investigations focus mainly on the static
properties of SO coupled BECs and have reveal rich phase diagrams
of the ground-states \cite{Wang,Jian,Ho,Congjun,Xu,Machida} and
exotic vortex structures \cite{Sarang,Santos,Hu,
Han,Congjun_2,Galitski_2}. However, to our knowledge, their dynamics has been less
studied \cite{Galitski,Merkl,Zhang,Zhu}, where the
SO coupled BECs are demonstrated to exhibit interesting
relativistic dynamics, such as analogs of self-localization
\cite{Merkl}, {\sl Zitterbewegung} \cite{Zhang} and Klein
tunneling \cite{Zhu} under certain conditions.

\begin{figure}[tbph]
\vspace{0.5cm}
\label{Fig1} 
\includegraphics[height=3.5cm,width=8cm]{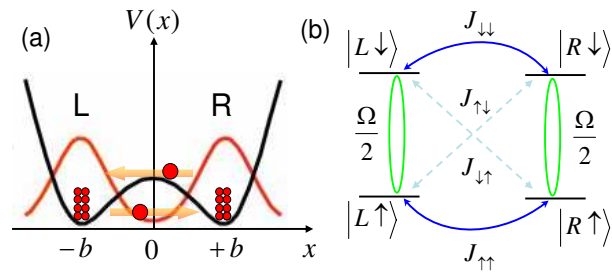}
\caption{(Color online) A schematic representation of (a) a SO
coupled BEC in a double well trap. (b) The dynamic process of the
system, where the blue solid and dashed lines represent the
inter-well tunneling without and with spin-flipping, respectively,
the green circles represent the Raman coupling. The inter-well
spin-flipping tunneling induced by the Raman coupling is negligible
under current experiment conditions \cite{NIST_SOC,Albiez}. The
atomic interaction terms are not shown in the figure.}
\end{figure}

On the other hand, quantum dynamics of a BEC in a double well
potential has been widely investigated. In particular, the
coherent atomic tunneling between two wells results in oscillatory
exchange of the BEC, which is analogous to the Josephson effects
(JEs) for neutral atoms
\cite{Javanaien,Milburn,Smerzi,Giovanazzi}. The weakly interacting
BECs provide a further context \cite{Milburn,Smerzi,Giovanazzi}
for JEs in superconductor systems because they display a nonlinear
generalization of typical d.c. and a.c. JEs and macroscopic
quantum self-trapping, all of which have been observed in
experiments \cite{Levy,Albiez,LeBlanc}. Apart from the
conventional single-species BECs
\cite{Javanaien,Milburn,Smerzi,Giovanazzi}, the Josephson dynamics
of two-species BECs \cite{Inguscio} and spinor BECs without SO
coupling \cite{Chang} have also been studied
\cite{YQLi,Pindzola,Pu,WMLiu}; however, the dynamics of SO coupled BECs
is yet to be explored.

In this paper, we investigate the dynamics of a specific SO coupled
BEC, which was realized in the experiment of the NIST group, in a
double well trapping potential. We find that the SO coupling in the
system contributes to and increases the atomic tunneling to a large
extent, which can significantly enhance the atomic  JEs. The full
dynamics of the system contains both internal and external JEs,
which are moreover inherently coupled in a complicated form even in
the absence of interatomic interactions. We further demonstrate that
the spin-dependent Josephson tunneling can lead to a net atomic spin
current by varying conditions, which we refer to as spin Josephson
effects. The predicted spin-Josephson currents are robust against
the parameter adjustment and varying initial conditions, and can be
observable in the SO coupled BECs under realistic experimental
conditions.

The paper is organized as follows. In Sec. II we construct a model
that can be used to study the quantum dynamics of a SO coupled BEC
in a double well potential. Then, in Sec. III, the Josephson
dynamics of the constructed system is investigated, with the
complicatedly coupled internal and external JEs being addressed. In
Sec. IV we demonstrate that the spin JEs exhibit in the system under
realistic conditions. Brief discussions and a short conclusion are
given in Sec. V.

\section{model}


In a very recent experiment, the NIST group realized a
synthetic SO coupling in the $^{87}$Rb BEC, in which a pair of
Raman lasers generate a momentum-sensitive coupling between two
internal atomic states \cite{NIST_SOC}. In the bare pseudo-spin
basis $|\uparrow\rangle_b=|m_F=0\rangle$ and
$|\downarrow\rangle_b=|m_F=-1\rangle$, the SO coupling is
described by the single particle Hamiltonian given by
\cite{NIST_SOC}
\begin{equation}
\label{NISTHam1}
\hat{h}=\frac{\mathbf{p}^2}{2m}\hat{I}+\frac{1}{2}\left(
                                                                \begin{array}{cc}
                                                                  \delta & \Omega e^{2ik_Lx} \\
                                                                  \Omega e^{-2ik_Lx} & -\delta \\
                                                                \end{array}
                                                              \right),
\end{equation}
 where $\mathbf{p}$ is the
atomic momentum in the $xy$ plane, $m$ is the atomic mass,
$\delta$ is the tunable detuning behaved as a Zeeman filed, $k_L$
is the wave number of the Raman laser, and $\Omega$ is the Raman
coupling strength.  Such kind of SO coupling is equivalent to that
of an electronic system with equal contribution from Rashba and
Dresselhaus SO coupling, and thus it is effective just in
one-dimension (1D). So we restrict our discussions in 1D and focus
on the motion of atoms along $x$ axis by freezing their $y$ and
$z$ degrees of freedom.

To proceed further, we introduce the dressed pseudo-spins
$|\uparrow\rangle=e^{-ik_Lx}|\uparrow\rangle_b$ and
$|\downarrow\rangle=e^{ik_Lx}|\downarrow\rangle_b$\cite{NIST_SOC,Galitski_2},
then the single-particle Hamiltonian in 1D (along $x$ axis) can be
written as
\begin{equation}
\label{NISTHam2}
\hat{h}_0=\frac{\hbar^2\hat{k}^2_x}{2m}+2\alpha\hat{k}_x\sigma_z+\frac{\Omega}{2}\sigma_x
+\frac{\delta}{2}\sigma_z,
\end{equation}
where $\hat{k}_x$ is the atomic wave vector operator, and
$\alpha=E_r/k_L$ is the SO coupling strength with
$E_r=\hbar^2k^2_L/2m$ being the single-photon recoil energy. The
dispersion relation of the single particle Hamiltonian
(\ref{NISTHam2}) with $\delta=0$ is
$E_{\pm}(k_x)=\frac{\hbar^2k_x^2}{2m}\pm
\sqrt{4\alpha^2k_x^2+\Omega^2/4}$, which exhibits a structure of two branches. We are interested in the lower energy one $E_{-}(k_x)$.
There is only one minimum in $k_x=0$ for large Raman coupling
$\Omega>4E_r$, where the atoms of both atomic levels condense.
However, the lower branch for $\Omega<4E_r$ presents two minima
for condensation of dressed pseudo-spin-up (left one) atoms and
dressed pseudo-spin-down (right one) atoms, respectively. The
Raman coupling and a small $\delta$ modulate the population of
atoms in these two states \cite{NIST_SOC}. Here we focus
on the later regime, i.e. $\Omega<4E_r$, because such BEC with
spin-separated and non-zero central momentum is more interesting
in contrast to a regular BEC with zero central momentum.

To be clearer, we can rewrite the Hamiltonian
(\ref{NISTHam2}) as
\begin{equation}
\hat{h}_0=\left(
          \begin{array}{cc}
            H_{\uparrow} & \Omega/2 \\
            \Omega/2 & H_{\downarrow} \\
          \end{array}
        \right),
\end{equation}
where
$H_{\uparrow}=\frac{\hbar^2}{2m}(\hat{k}^2_x+2k_L\hat{k}_x)+\frac{\delta}{2}$
and
$H_{\downarrow}=\frac{\hbar^2}{2m}(\hat{k}^2_x-2k_L\hat{k}_x)-\frac{\delta}{2}$.
Since it is  more straightforward to describe the system in terms of
dressed pseudo-spin states compared with using bare ones, we will
work in the dressed pseudo-spin space and simply refer to dressed
pseudo-spin as spin for convenience hereafter. We also note that the
parameters $k_L$, $\Omega$ and $\delta$ in the single-particle
Hamiltonian can be tuned independently in a wide range
\cite{NIST_SOC}, making the SO coupled BEC a suitable platform for
investigating the Josephson dynamics in the presence of SO coupling.


Now we turn to consider such a SO coupled BEC in a double well
potential denoted by $V(x)$ as shown in Fig. 1(a). Note that the
double well potential here is assumed to be spin-independent. To
investigate the dynamics of the system, we adopt the two-mode
approximation \cite{Javanaien,Milburn,Smerzi,Giovanazzi} with the
field operator
\begin{equation}
\label{FieldOp}
\hat{\Psi}_{\sigma}(x)\simeq\hat{a}_{L\sigma}\psi_{L\sigma}(x)+\hat{a}_{R\sigma}\psi_{R\sigma}(x),
\end{equation}
where $\psi_{j\sigma}(x)$ is the ground state wave function of the
$j$ well ($j=L,R$)  with spin $\sigma$
($\sigma=\uparrow,\downarrow$), and $\hat{a}_{j\sigma}$ is the
annihilation operator for spin $\sigma$ in the $j$ well,
satisfying the bosonic commutation relationship
$[\hat{a}_{j\sigma},\hat{a}^{\dag}_{k\sigma'}]=\delta_{jk}\delta_{\sigma\sigma'}$.
The validity of the two-mode approximation holds under two
conditions: the weak atomic interaction and small effective Zeeman
splitting, as the atoms can not be pumped out of the lowest
state of each well in this case. In the second quantization formalism, the total
Hamiltonian reads
\begin{equation}
\label{FullHam1} \mathcal{H}=\int dx
\hat{\Psi}^{\dag}(x)\left[\hat{h}_0+V(x)+\hat{h}_{int}
\right]\hat{\Psi}(x),
\end{equation}
where the two-component field operator
$\hat{\Psi}=(\hat{\Psi}_{\uparrow},\hat{\Psi}_{\downarrow})^{\text{T}}$,
and the interaction Hamiltonian $\hat{h}_{int}$ will be specified
 below. Substituting Eq. (\ref{FieldOp}) into Eq.
(\ref{FullHam1}), one can rewrite the total Hamiltonian as
%
%
\begin{equation}
\begin{array}{ll}
\label{FullHam2} \displaystyle \mathcal{H}=\sum_{j,\sigma}
\varepsilon_{j\sigma} \hat{a}^{\dag}_{j\sigma}\hat{a}_{j\sigma} +
\sum_{\sigma\sigma'}\left(J_{\sigma\sigma'}\hat{a}^{\dag}_{L\sigma}\hat{a}_{R\sigma'}+h.c.
\right) \\ \qquad+ \displaystyle
\frac{\Omega}{2}\sum_{j}\left(\hat{a}^{\dag}_{j\uparrow}\hat{a}_{j\downarrow}+h.c.
\right)\\ \qquad+ \displaystyle
\frac{\delta}{2}\sum_{j}\left(\hat{a}^{\dag}_{j\uparrow}\hat{a}_{j\uparrow}-\hat{a}^{\dag}_{j\downarrow}\hat{a}_{j\downarrow}
\right)+\mathcal{H}_{int},
\end{array}
\end{equation}
where $\varepsilon_{j\sigma}=\int dx \psi^{\ast}_{j\sigma}(x)
[H_{\sigma}+V(x)] \psi_{j\sigma}(x)\approx
\frac{1}{2}\hbar\omega_j-E_r$ is the single-particle ground state
energy in the $j$ well with $\omega_j$ being the harmonic
frequency of this well,
$J_{\sigma\sigma}=\int dx
\psi^{\ast}_{L\sigma}(x) [H_{\sigma}+V(x)] \psi_{R\sigma}(x)$ and
$J_{\sigma\bar{\sigma}}=\int dx \psi^{\ast}_{L\sigma}(x)
\frac{\Omega}{2} \psi_{R\bar{\sigma}}(x)$ (with $\sigma$ and
$\bar{\sigma}$ referring to different spins) are the tunneling
terms shown in Fig. 1(b).  In addition,
the interaction Hamiltonian is given by
\begin{equation}
\begin{array}{ll}
\mathcal{H}_{int} =
\displaystyle\frac{1}{2}\sum_{j}\left(g_{\uparrow\uparrow}^{(j)}\hat{a}^{\dag}_{j\uparrow}
\hat{a}^{\dag}_{j\uparrow} \hat{a}_{j\uparrow} \hat{a}_{j\uparrow} +
g_{\downarrow\downarrow}^{(j)}\hat{a}^{\dag}_{j\downarrow}
\hat{a}^{\dag}_{j\downarrow} \hat{a}_{j\downarrow}
\hat{a}_{j\downarrow}\right.\nonumber\\
\left.\qquad\qquad\qquad+
2g_{\uparrow\downarrow}^{(j)}\hat{a}^{\dag}_{j\uparrow}
\hat{a}^{\dag}_{j\uparrow} \hat{a}_{j\downarrow}
\hat{a}_{j\downarrow} \right),
\end{array}
\end{equation}
where
$g_{\sigma\sigma'}^{(j)}=\frac{2\hbar^2a_{\sigma\sigma'}}{ml_\bot^2}
\int dx |\psi_{j\sigma}(x)|^2 |\psi_{j\sigma'}(x)|^2$ is the effective 1D interacting
strength with
$a_{\sigma\sigma'}$ being the $s$-wave scattering length between
spin $\sigma$ and $\sigma'$ and $l_\bot$ being the oscillator
length associated to a harmonic vertical confinement.
 Note that here we have
ignored the inter-well atomic interactions because the $s$-wave
scattering length (which is on the order of nanometers) is much
smaller than the inter-well distance (which is on the order of
micrometers). We have also dropped the inter-well coupling since
its strength is exponentially smaller than the intra-well
counterpart. The Hamiltonian (\ref{FullHam2}) describes the
dynamic process of the system schematically shown in Fig. 1(b).

\begin{figure}[tbph]
\vspace{0.5cm}
\label{Fig2} 
\includegraphics[height=6.5cm,width=8cm]{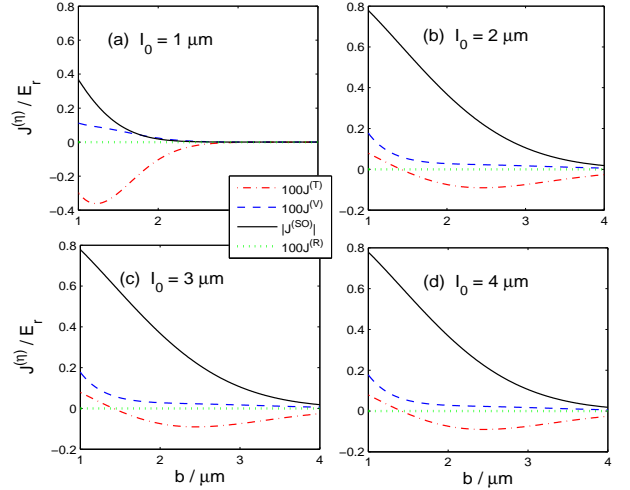}
\caption{(Color online) The energy scales of tunneling terms
$J^{(\eta)}$ as a function of $b$ for (a) $l_0=1$ $\mu$m, (b)
$l_0=2$ $\mu$m, (c) $l_0=3$ $\mu$m, and (d) $l_0=4$ $\mu$m,
respectively. In (a)-(d) we set $\Omega/\hbar=E_r/\hbar=22.5$ kHz.
}
\end{figure}

For simplicity, we assume the double well potential to be symmetric
as shown in Fig. 1(a), with each well having the same harmonic
trapping frequency $\omega$. Thus we have
$\varepsilon_L=\varepsilon_R$ and
$g_{\sigma\sigma'}^{(L)}=g_{\sigma\sigma'}^{(R)}$. Such kind of
double well potential can be generated in experiments \cite
{LeBlanc} with the form
\begin{equation}
\label{Doublewell} V(x)=a(x^2-b^2)^2,
\end{equation}
where the parameters $a$ and $b$ are both tunable in the
experiments \cite{LeBlanc}. Expanding $V(x)$ near $x=\pm b$, one
obtains its harmonic form as
$V^{\text{(2)}}(x)\doteq\frac{1}{2}m\omega^2(x\pm b)^2$ and thus
$a=m\omega^2/8b^2$. The ground state wavefunctions of the BEC in
each well potential with each spin can be approximately
represented by its corresponding lowest energy single-particle
wavefunction, which can be worked out by solving the equations
$[H_{\sigma}+\frac{1}{2}m\omega^2(x\pm
b)^2]\psi_{j\sigma}=\varepsilon_{j\sigma}\psi_{j\sigma}$ (here
$\pm$ are for $j=L,R$, respectively). The results are \cite{Galitski}
\begin{equation}
\label{groundstate}
\begin{array}{ll}
\psi_{L\uparrow} = \varphi_0^{(L)}(x) e^{-ik_Lx},\\
\psi_{L\downarrow} = \varphi_0^{(L)}(x) e^{ik_Lx},\\
\psi_{R\uparrow} = \varphi_0^{(R)}(x) e^{-ik_Lx},\\
\psi_{R\downarrow} = \varphi_0^{(R)}(x) e^{ik_Lx},\\
\end{array}
\end{equation}
where
$\varphi_0^{(L)}(x)=\frac{1}{\sqrt{l_0\sqrt{\pi}}}e^{-(x+d)^2/2l_0^2}$,
and
$\varphi_0^{(R)}(x)=\frac{1}{\sqrt{l_0\sqrt{\pi}}}e^{-(x-d)^2/2l_0^2}$
with $l_0=\sqrt{\hbar/m\omega}$ being the oscillator length.
Substituting Eq. (\ref{groundstate}) into the expressions of
$J_{\sigma\sigma'}$, one can obtain
\begin{equation}
\label{tunnelingenergy1}
\begin{array}{ll}
J_{\uparrow\uparrow} = J^{(T)}+J^{(SO)}+J^{(V)}+J^{(Z)},\\
J_{\downarrow\downarrow} = J^{(T)}+J^{(SO)}+J^{(V)}-J^{(Z)},\\
J_{\uparrow\downarrow} = J_{\downarrow\uparrow}=J^{(R)},
\end{array}
\end{equation}
where the terms $J^{(T)}=-\frac{\hbar^2}{2m}\int dx \varphi_0^{(L)}
\varphi_0^{''(R)}$, $J^{(SO)}=-E_r\int dx \varphi_0^{(L)}
\varphi_0^{(R)}$, $J^{(V)}=\int dx \varphi_0^{(L)} V(x)
\varphi_0^{(R)}$, $J^{(Z)}=\frac{\delta}{2}\int dx \varphi_0^{(L)}
\varphi_0^{(R)}$, and $J^{(R)}=\frac{\Omega}{2}\int dx
\varphi_0^{(L)} e^{-2ik_Lx} \varphi_0^{(R)}$.

Compared with the atomic tunneling of a regular BEC, the SO coupled
BEC in this system exhibits two additional tunneling channels, the
SO coupling induced tunneling term $J^{(SO)}$ and the Raman coupling
induced one $J^{(R)}$. To clarify the effects of these terms in the
tunneling processes, we need to work out and to compare the energy
scales of all the terms $J^{(\eta)}$, where $\eta = \{T, V, Z, SO,
R\}$. Substituting Eqs. (\ref{groundstate}) and (\ref{Doublewell})
into Eq. (\ref{tunnelingenergy1}), we can obtain the following
analytical solutions
\begin{equation}
\label{tunnelingenergy2}
\begin{array}{ll}
J^{(\eta)}=\xi_{\eta} e^{-b^2/l^2_0},
\end{array}
\end{equation}
where
$\xi_T=\frac{\hbar^2}{2ml_0^2}(\frac{1}{2}-\frac{b^2}{l^2_0})$,
$\xi_V=\frac{\hbar^2}{8mb^2l^4_0}(\frac{3}{4}l^4_0-b^2l^2_0+b^4)$,
$\xi_Z = \delta/2 $, $\xi_{SO}=-E_r$, and $\xi_{R}=\Omega
e^{-k_L^2l_0^2}$. Since the Zeeman filed $\delta$ is independently
tunable to the double-well structure and should be small, we here
further assume $\delta\ll E_r$ and thus we focus on the comparison
among $J^{(SO, R, T, V)}$. The effects of Zeeman-splitting induced
tunneling will be specified in the Sec. IV.

For $\l_0\sim\sqrt{2}b$, we have $\xi_T\sim0$ and $\xi_V\sim
\frac{\hbar^2}{4mb^2}$, and thus
$\frac{|\xi_{SO}|}{|\xi_V|}=8\pi^2(\frac{b}{\lambda_L})^2\gtrsim100$.
Here we have assumed the same wavelength $\lambda_L=2\pi/k_L=0.8$
$\mu$m and recoil frequency $E_r/\hbar=22.5$ kHz as those in the
experiments \cite{NIST_SOC}, $b$ and $l_0$ to be on the order of
micrometers \cite{Albiez,LeBlanc}. In fact, in the regime of
$b^2/l^2_0\sim[0.5,2]$, we find that
\begin{equation}
|\xi_{SO}|\gtrsim100 \max\{|\xi_{T}|,|\xi_{V}|\}. \end{equation}
Besides, one can check that $\xi_{R}\sim\Omega e^{-64}$ for
$\l_0\sim1\mu$m and $\lambda_L=0.8\mu$m, and thus the
Raman-coupling induced tunneling is negligible in this system. The
comparisons among $J^{(\eta)}$ for some typical parameters are
shown in Fig. 2. In other words, we find that under realistic
experiment conditions \cite{NIST_SOC,Albiez}, the spin-flipping
tunneling induced by Raman coupling is negligible but the SO
coupling induced tunneling term $J^{(SO)}$ dominates and moreover
it greatly enhances atomic tunneling in this system. Thus we may
rewrite the tunneling terms as
\begin{equation}
\label{tunnelingenergy3}
\begin{array}{ll}
J_{\uparrow\downarrow}=J_{\downarrow\uparrow} \approx 0, \\
J_{\uparrow\uparrow}\approx J_{\downarrow\downarrow} \approx J^{(SO)}=-\gamma E_r,\\
\end{array}
\end{equation}
where $\gamma=\exp(-b^2/l_0^2)\sim[0.1,0.6]$. It is worthwhile to
note that the new tunneling terms $J^{(SO)}$ and  $J^{(R)}$ in
this SO coupled system are both tunable, enabling us to
study the interesting effects of SO coupling in the atomic
inter-well tunneling. For instance, one can decrease the effective
wave number in $x$ axis to the scale $k_L\sim1/l_0$ so that
$J^{(R)}\sim0.37\Omega$ and then the Raman-coupling induced
tunneling can revive. This can be achieved by adjusting the angle
between the applying Raman lasers and the trapping potential or
alternatively by using lasers with larger wavelength. In addition,
in the same way one can tune the recoil energy $E_r$ to identify the
enhancement of atomic tunneling due to the SO coupling (i.e. the
effect of $J^{(SO)}$) in experiments. As a first step to investigate
the system under current experiment conditions, we here concentrate on the
tunneling regime governed by Eq. (\ref{tunnelingenergy3}).


\section {Full dynamics of the system}

We are now in the position to investigate the quantum dynamics of
the system constructed in the previous section.  We first address
the non-interacting case, i.e. $\mathcal{H}_{int}=0$ in Eq.
(\ref{FullHam2}), in which the single-particle Hamiltonian is given
by
\begin{equation}
\label{SingleHam}
\begin{array}{ll}
\mathcal{H}_0 \simeq
J_{\uparrow\uparrow}\left(\hat{a}^{\dag}_{L\uparrow}\hat{a}_{R\uparrow}
+ \hat{a}_{L\uparrow}\hat{a}^{\dag}_{R\uparrow} \right) +
J_{\downarrow\downarrow}\left(\hat{a}^{\dag}_{L\downarrow}\hat{a}_{R\downarrow}
+ \hat{a}_{L\downarrow}\hat{a}^{\dag}_{R\downarrow} \right)\\
\qquad+\frac{\Omega}{2}\left(\hat{a}^{\dag}_{L\uparrow}\hat{a}_{L\downarrow}
+ \hat{a}_{L\uparrow}\hat{a}^{\dag}_{L\downarrow} +
\hat{a}^{\dag}_{R\uparrow}\hat{a}_{R\downarrow} +
\hat{a}_{R\uparrow}\hat{a}^{\dag}_{R\downarrow}\right)\\
\qquad+\frac{\delta}{2}\left(\hat{a}^{\dag}_{L\uparrow}\hat{a}_{L\uparrow}-
\hat{a}^{\dag}_{L\downarrow}\hat{a}_{L\downarrow}+
\hat{a}^{\dag}_{R\uparrow}\hat{a}_{R\uparrow} -
\hat{a}^{\dag}_{R\downarrow}\hat{a}_{R\downarrow}\right).
\end{array}
\end{equation}
Here we have dropped the tunneling terms $J_{\sigma\bar{\sigma}}$
since these spin-flipping tunneling progresses can be negligible
in the current experiment conditions \cite{NIST_SOC,Albiez}. In
order to study the dynamic properties of the system, we need to
work with the equation of motion.
The corresponding Heisenberg equations read
\begin{equation}
\label{Heisenberg}
\begin{array}{ll}
i\hbar\frac{d}{dt} \hat{a}_{j\sigma} =
\left[\hat{a}_{j\sigma},\mathcal{H}_0 \right]=
J_{\sigma\sigma}\hat{a}_{\bar{j}\sigma} +
\frac{\Omega}{2}\hat{a}_{j\bar{\sigma}}+(-1)^p\frac{\delta}{2}\hat{a}_{j\sigma},
\end{array}
\end{equation}
where $\sigma$ and $\bar{\sigma}$ refer to different spin, while $j$
and $\bar{j}$ to different wells, and $p=0,1$ are for
$\sigma=\uparrow,\downarrow$, respectively. Using the mean-field
approximation, one has
$\hat{a}_{j\sigma}\simeq\langle\hat{a}_{j\sigma}\rangle\equiv
a_{j\sigma}$ with $a_{j\sigma}$ being $c$ numbers. Thus we can
rewrite the equations of motion as
\begin{equation}
\label{MotionEqCN}
\begin{array}{ll}
i\hbar \dot{a}_{j\sigma} = J_{\sigma\sigma}a_{\bar{j}\sigma} +
\frac{\Omega}{2}a_{j\bar{\sigma}}+(-1)^p\frac{\delta}{2}a_{j\sigma}.
\end{array}
\end{equation}
By defining a four-component wavefunction $\Phi=\left(
a_{L\uparrow},a_{L\downarrow},a_{R\uparrow},a_{R\downarrow}\right)^{\text{T}}$,
 Eq. (\ref{MotionEqCN}) is rewritten as
$i\hbar\frac{d}{dt}\Phi=H_M\Phi$, where the Hamiltonian of the
system is given by
\begin{equation}
\label{MotionHam} H_M =\left(
                         \begin{array}{cccc}
                           \frac{\delta}{2} & \frac{\Omega}{2} & J_{\uparrow\uparrow} & 0 \\
                           \frac{\Omega}{2} & -\frac{\delta}{2} & 0 & J_{\downarrow\downarrow} \\
                           J_{\uparrow\uparrow} & 0 & \frac{\delta}{2} & \frac{\Omega}{2} \\
                           0 & J_{\downarrow\downarrow} & \frac{\Omega}{2} & -\frac{\delta}{2} \\
                         \end{array}
                       \right).
\end{equation}

We now look into the JEs in this system. Let us further express
$a_{j\sigma}$ as
$a_{j\sigma}=\sqrt{N_{j\sigma}}e^{i\theta_{j\sigma}}$, where the
particle numbers $N_{j\sigma}$ and phases $\theta_{j\sigma}$ are
all time-dependent in general. According to Eq.
(\ref{MotionEqCN}), we can obtain
\begin{equation}
\begin{array}{ll}
\label{MotionFurther} i\hbar\frac{\dot{N}_{j\sigma}}{2}-\hbar
N_{j\sigma} \dot{\theta}_{j\sigma} =
J_{\sigma\sigma}\sqrt{N_{j\sigma}N_{\bar{j}\sigma}}e^{i(\theta_{\bar{j}\sigma}-\theta_{j\sigma})}\\
\qquad\qquad+\frac{\Omega}{2}\sqrt{N_{j\sigma}N_{j\bar{\sigma}}}e^{i(\theta_{j\bar{\sigma}}-\theta_{j\sigma})}+(-1)^p\frac{\delta}{2}N_{j\sigma}.
\end{array}
\end{equation}
Separating the image and real parts of Eq.
(\ref{MotionFurther}) yields two groups of equations as
\begin{widetext}
\begin{equation}
\label{MotionTwoPart}
\begin{array}{ll}
\dot{N}_{j\sigma} =
\frac{2J_{\sigma\sigma}}{\hbar}\sqrt{N_{j\sigma}N_{\bar{j}\sigma}}\sin(\theta_{\bar{j}\sigma}-\theta_{j\sigma})
+\frac{\Omega}{\hbar}\sqrt{N_{j\sigma}N_{j\bar{\sigma}}}\sin(\theta_{j\bar{\sigma}}-\theta_{j\sigma}),\\
\dot{\theta}_{j\sigma} =
\frac{J_{\sigma\sigma}}{\hbar}\sqrt{\frac{N_{\bar{j}\sigma}}{N_{j\sigma}}}\cos(\theta_{\bar{j}\sigma}-\theta_{j\sigma})
+\frac{\Omega}{2\hbar}\sqrt{\frac{N_{j\bar{\sigma}}}{N_{j\sigma}}}
\cos(\theta_{j\bar{\sigma}}-\theta_{j\sigma})+(-1)^p\frac{\delta}{2\hbar}.
\end{array}
\end{equation}
\end{widetext}
Eq. (\ref{MotionTwoPart}) consists actually of eight coupled
equations. To simplify these equations, we introduce
$\phi_{\sigma}=\theta_{R\sigma}-\theta_{L\sigma}$ and
$\rho_{\sigma}=N_{R\sigma}-N_{L\sigma}$ for the phase and particle
number differences between two wells with the same spin $\sigma$,
and $\phi_{j}=\theta_{j\downarrow}-\theta_{j\uparrow}$ and
$\rho_{j}=N_{j\downarrow}-N_{j\uparrow}$ for the phase and particle
number differences between two spins in the same well $j$,
respectively. Thus we can obtain
\begin{eqnarray}
 \dot{\rho}_{\sigma} &=&    \mathcal{L}_1\sin\phi_\sigma+ \frac{1}{2}\sum_j
(-1)^q\mathcal{L}_2 \sin\phi_j, \nonumber \\
\label{Josephsoneffect}
 \dot{\rho}_{j}&=& \frac{1}{2}\sum_{\sigma}(-1)^p
\mathcal{L}_1 \sin\phi_{\sigma} + \mathcal{L}_2 \sin\phi_j,
 \end{eqnarray}
%
%
where
$\mathcal{L}_1=-\frac{4J_{\sigma\sigma}}{\hbar}\sqrt{N_{R\sigma}N_{L\sigma}}$,
$\mathcal{L}_2=-\frac{2\Omega}{\hbar}\sqrt{N_{j\uparrow}N_{j\downarrow}}$,
and $q=0,1$ for $j=R,L$, respectively ($p=0,1$ for
$\sigma=\uparrow,\downarrow$, respectively). From the above Eq.
(\ref{Josephsoneffect}), we find the coexistence of internal JE
related to $\dot{\rho}_{j}(\phi_{j})$ and external JE related to
$\dot{\rho}_{\sigma}(\phi_{\sigma})$. Moreover, the internal and
external JEs are inherently coupled in a more complicated form.

Before ending this section, we briefly discuss the weakly
interacting cases, which have been assumed to meet the requirement
of two-mode approximation. In this regime, the mean-field analysis
still works well, and the dropped term $\mathcal{H}_{int}$  can be
taken into count within the previous discussions. This leads to two
additional terms related to interactions into Eq.
(\ref{MotionEqCN}), and now the equations of motion are given by
\begin{equation}
\begin{array}{ll}
\label{MotionEqINT} i\hbar \dot{a}_{j\sigma} =
J_{\sigma\sigma}a_{\bar{j}\sigma} +
\frac{\Omega}{2}a_{j\bar{\sigma}}+(-1)^p\frac{\delta}{2}a_{j\sigma}\\
\qquad\qquad+g_{\sigma\sigma}|a_{j\sigma}|^2a_{j\sigma}+g_{\sigma\bar{\sigma}}|a_{j\bar{\sigma}}|^2a_{j\sigma},
\end{array}
\end{equation}
where the interacting strength $g_{\sigma\sigma'}$ can be found
as
$g_{\sigma\sigma'}=\frac{\sqrt{2}\hbar^2a_{\sigma\sigma'}}{\sqrt{\pi}ml_{\bot}^2l_0}$.
The estimation of the interaction energy and the Josephson dynamics
in the presence of weak interactions will be presented in the next
section.

\section{Josephson effects in weak Raman coupling regimes}

In the preceding section, we have shown that the SO coupled BEC in a
double well potential exhibits the complicated coupled external and
internal Josephson dynamics. We, in this section, consider a
specific dynamic process of the system in the weak Raman coupling
regime (i.e. $\Omega/E_r\ll 1$), where the external Josephson
dynamic dominates. In fact, the manipulation and detection of the SO
coupled BECs in this regime have been performed in experiments
\cite{NIST_SOC}.

For the weak Raman coupling,  we find that the ratio $\nu\equiv
|J_{\sigma\sigma}|/\Omega$ can reach several hundreds from Eq.
(\ref{tunnelingenergy3}). Thus within the time scale
$\tau\sim\hbar/\Omega\simeq45$ ms for $\Omega= 0.001E_r$, one can
ignore the effects of the spin-flipping tunneling, which leads to
two external Josephson tunneling processes for different spins. The
spins in this regime are conserved and then the total particle
number of spin $\sigma$ $N_{\sigma t}=N_{L\sigma}+N_{R\sigma}$ are
time-independent constants. We assume $N_{\uparrow t}=N_{\downarrow
t}=N_t$ for simplicity. The equations of motion (\ref{MotionEqINT})
in this case can be rewritten as
\begin{widetext}
\begin{equation}
\label{MotionSpin} i\hbar\frac{d}{dt}\left(
                                       \begin{array}{c}
                                         a_{L\sigma} \\
                                         a_{R\sigma} \\
                                       \end{array}
                                     \right)=\left(
                                               \begin{array}{cc}
                                                 (-1)^p\frac{\delta}{2}+g_{\sigma\sigma}|a_{L\sigma}|^2+g_{\sigma\bar{\sigma}}|a_{L\bar{\sigma}}|^2 & J_{\sigma\sigma}\\
                                                 J_{\sigma\sigma} & (-1)^p\frac{\delta}{2}+g_{\sigma\sigma}|a_{R\sigma}|^2+g_{\sigma\bar{\sigma}}|a_{R\bar{\sigma}}|^2 \\
                                               \end{array}
                                             \right)\left(
                                       \begin{array}{c}
                                         a_{L\sigma} \\
                                         a_{R\sigma} \\
                                       \end{array}
                                     \right).
\end{equation}
\end{widetext}
By defining the normalized inter-well particle number difference for
spin $\sigma$ as
$\mathcal{Z}_{\sigma}=[N_{R\sigma}-N_{L\sigma}]/N_t$ ($-1 \le
\mathcal{Z}_{\sigma} \le 1 $), the equations
of motion (\ref{MotionTwoPart}) become rather simple in this case
(similar to those for the regular two species BECs \cite{YQLi}),
which are given by
\begin{equation}
\begin{array}{ll}
\label{Timeevolution} \dot{\mathcal{Z}}_{\sigma} =
-\frac{2J_{\sigma\sigma}}{\hbar}\sqrt{1-\mathcal{Z}_{\sigma}}\sin\phi_{\sigma},\\
\dot{\phi}_{\sigma}=\frac{J_{\sigma\sigma}}{\hbar}\frac{\mathcal{Z}_{\sigma}}{\sqrt{1-\mathcal{Z}^2_{\sigma}}}\cos\phi_{\sigma}
+\frac{U_{\sigma\sigma}}{\hbar}\mathcal{Z}_{\sigma}+\frac{U_{\sigma\bar{\sigma}}}{\hbar}\mathcal{Z}_{\bar{\sigma}}+(-1)^p\frac{\delta}{2\hbar}.
\end{array}
\end{equation}
The spin-dependent atomic density current is given by
\begin{equation}
\label{Spincurrent} I_{\sigma}=N_t\cdot
\dot{\mathcal{Z}}_{\sigma}.
\end{equation}
From Eq.(\ref{Spincurrent}), we can define  the net spin current
as \begin{equation} I_s=I_{\uparrow}-I_{\downarrow},\end{equation}
and the total atomic current as
\begin{equation} I_a=I_{\uparrow}+I_{\downarrow}.\end{equation}

\begin{figure}[tbph]
\vspace{0.5cm}
\label{Fig2} 
\includegraphics[height=7cm,width=8cm]{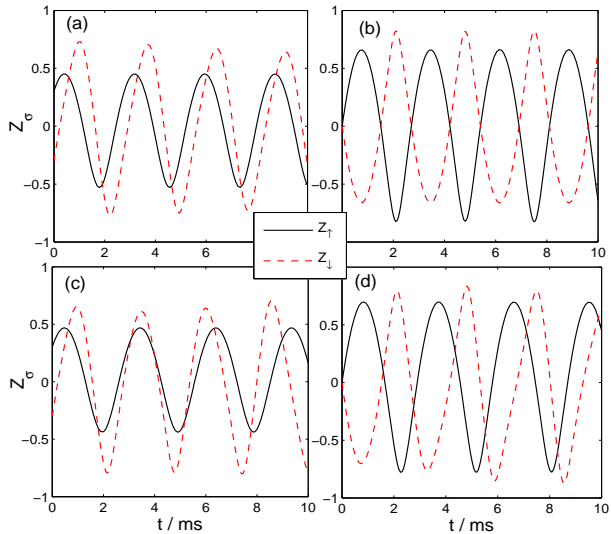}
\caption{(Color online) The time evolution of $\mathcal{Z}_{\sigma}$
in noninteracting limitation. In (a) and (b) we have $\delta=0$ and
$J_{\uparrow\uparrow}=J_{\downarrow\downarrow}=-0.1E_r$; In (c) and
(d) we have $\delta=0.01E_r$, $J_{\uparrow\uparrow}=-0.905E_r$, and
$J_{\downarrow\downarrow}=-0.105E_r$. The initial conditions are
$\mathcal{Z}_{\uparrow}(0)=-\mathcal{Z}_{\downarrow}(0)=0.3$,
$\phi_{\uparrow}(0)=0.5\phi_{\downarrow}(0)=\pi/4$ in (a) and (c);
and $\mathcal{Z}_{\uparrow}(0)=\mathcal{Z}_{\downarrow}(0)=0$,
$\phi_{\uparrow}(0)=\phi_{\downarrow}(0)=\pi/4$ in (b) and (d).}
\end{figure}

\begin{figure}[tbph]
\vspace{0.5cm}
\label{Fig2} 
\includegraphics[height=6cm,width=8cm]{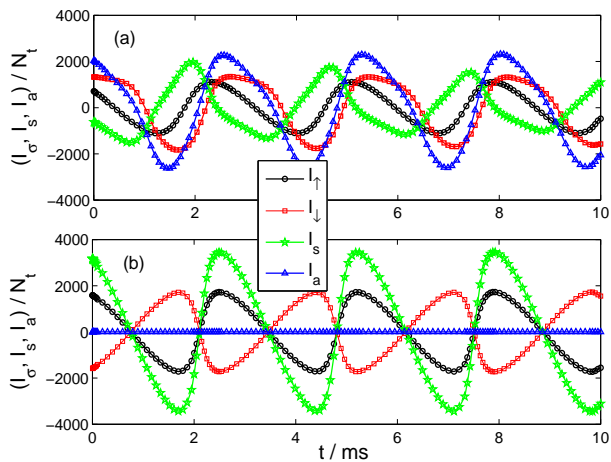}
\caption{(color online)  Josephson currents in the noninteracting limit.
The time evolution of spin-dependent atomic currents $I_{\sigma}$, a
net spin current $I_s$ and a total atomic current $I_a$ in (a) for
the same conditions in Fig. 3(a); and in (b) for the some conditions
in Fig. 3(b).}
\end{figure}

We first consider the JEs of the system in the noninteracting limit,
i.e. $U_{\sigma\sigma}=U_{\sigma\bar{\sigma}}=0$ in Eq.
(\ref{Timeevolution}), which can be realized by Feshbach resonance
\cite{Chin}. Under this condition, the two external Josephson
tunneling processes for different spins are decoupled. We
numerically calculate Eqs. (\ref{Timeevolution}), with some typical
results of the time evolution of $\mathcal{Z}_{\sigma}$ for
different initial conditions being shown in Fig. 3. In the
calculations, we have assumed the zero Zeeman filed $\delta=0$ in
Fig. 3(a) and (b), and small Zeeman field $\delta=0.01E_r$ in (c)
and (d). Compared with zero Zeeman field cases, a small Zeeman field
results in a deviation in Josephson tunneling strengths
$J_{\sigma\sigma}$ and in time-cumulative phases $\delta/2\hbar$ for
different spins. Here $\mathcal{Z}_{\uparrow}(t)$ and
$\mathcal{Z}_{\downarrow}(t)$ demonstrate the oscillatory Josephson
tunnelings which are similar to the early results in Ref.
\cite{Smerzi}. As shown in Fig. 3(a-d), they are spin-dependent and
the dynamic evolution of each one depends on its own tunneling
strength, phase and initial conditions. We also calculate
$I_{\sigma}$, $I_s$ and $I_a$ in this regime with typical results
being shown in Fig. 4. It is interesting to see that the
spin-dependent atomic density currents due to the spin-related
Josephson tunnelings give rise to a net spin current (cf. Fig. 4),
and moreover in some certain initial conditions the total atomic
current can be zero, which leads to a new interesting pure spin
currents (cf. Fig. 4(b)). We call such new JEs as spin Josephson
effects, which can be observable in experiments by measuring the
time-evolution of spin-dependent population imbalance of the atomic
gas \cite{YAChen}.

\begin{figure}[tbph]
\vspace{0.5cm}
\label{Fig2} 
\includegraphics[height=7cm,width=8cm]{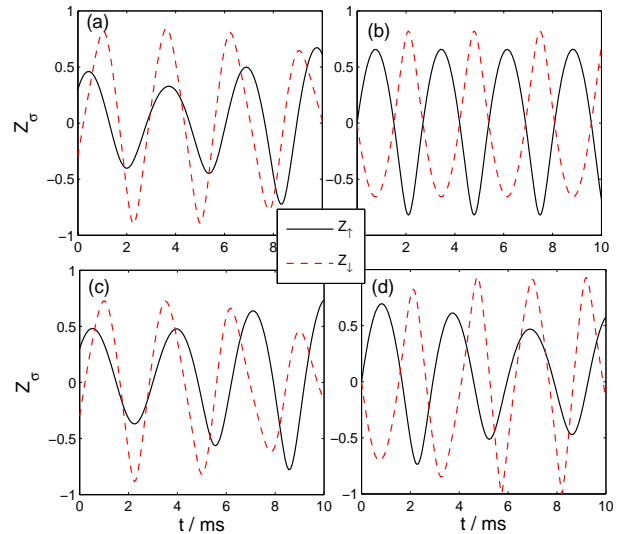}
\caption{(Color online) The time evolution of $\mathcal{Z}_{\sigma}$
in the weakly interacting regime with $U_{\sigma\sigma}=0.01E_r$ and
$U_{\sigma\sigma'}=0.011E_r$. Other parameters and initial
conditions in (a)-(d) are the same with those in Fig. 3(a)-(d),
respectively.}
\end{figure}

For weakly interacting cases, we have to estimate the interaction
energy $U_{\sigma\sigma}$ and $U_{\sigma\bar{\sigma}}$, which should
be $U_{\sigma\sigma}, U_{\sigma\bar{\sigma}} \ll \hbar\omega$ due to
the two-mode approximation. This requirement results in
$U_{\sigma\sigma}\approx U_{\sigma\bar{\sigma}}\ll0.1E_r$. In this
regime, the two spin-Josephson tunneling processes are coupled via
atomic interactions. To understand the effects of the interaction,
we show in Fig. 5 some typical results of the time evolution of
$\mathcal{Z}_{\sigma}$ for the same initial conditions and
parameters in Fig. 3. It clearly demonstrate that the modification
of $\mathcal{Z}_{\sigma}(t)$ due to atomic interactions is not
significant and even very minor in some cases (such as the case for
$\mathcal{Z}_{\sigma}(0)=0$ and $\delta=0$ in Fig. 3(b), 5(b)) since
the interaction energy is small compared with the tunneling energy.
Therefore the spin Josephson dynamics still exhibit a similar
oscillatory feature in this regime.

To see more clearly the oscillatory  properties of the spin JEs, we
have numerically calculated the frequency spectra of the net spin
currents $I_s$ for various conditions (such as those in Fig. 3 and
Fig. 5). We find that the spectra for different cases exhibit a
single peak centered at the slightly shifted frequency, as seen in
Fig. 6. The single-peak feature shown in Fig. 6 implies that the
spin current $I_s(t)$ can well be described by a $sin$-function,
while the weak interatomic interactions or the small Zeeman field
can merely modify the period and amplitude of the current slightly.
Thus we conclude that the spin JEs in this system are robust against
the parameter adjustment and initial conditions.

\begin{figure}[tbph]
\vspace{0.5cm}
\includegraphics[height=6cm,width=6.5cm]{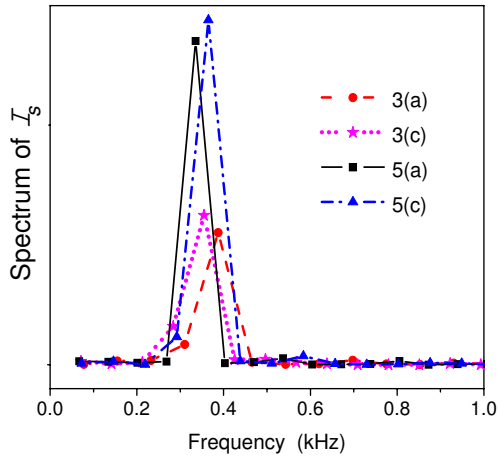}
\caption{(Color online)  Spectra of the net spin currents $I_s$
for the cases in Fig. 3(a,c) and in Fig. 5 (a,c). A single peak in
the spectrum of each case implies that the spin current
$I_s(t)$ is well described by a $sin$-function.}
\end{figure}

\section{Discussion and Conclusion}

Before concluding, we briefly discuss another specific dynamic
process of the system in the relatively strong Raman coupling regime,
$\Omega\gg |J_{\sigma\sigma}|$, which can be realized such as by
tuning $J_{\sigma\sigma}\sim-0.1E_r$ and $\Omega\sim E_r$. In this
regime, within the time scale $\hbar/|J_{\sigma\sigma}|$ one can
ignore the atomic inter-well tunneling and consider only the
internal dynamics in each single well. The atomic tunneling between
two spins refers to spin-flipping is induced by the Raman coupling,
and thus such Josephson tunneling is in the spin space. Considering
the atomic gas condenses with a finite but opposite momentum for
different spins, the internal JEs connect interesting quantum
tunneling in the momentum space.

In summary, we have investigated the quantum dynamics of a SO
coupled BEC in a symmetric double well potential. The SO coupling
contributes to atomic tunneling between wells and significantly
enhances JEs for realistic conditions. We have predicted a novel
spin Josephson effects which can be observed in a practical
experiment since all the required ingredients, including the SO
coupled BECs and the tunable double well potential, have already
been achieved in the previous experiments.

\vspace{1.0cm}

\acknowledgements This work was supported by the NBRPC
(No.2011CBA00302), the SKPBRC (Nos.2011CB922104 and 2011CB921503), the NSF of China
(Grant Nos. 10974059, 11075020 and 11125417), and the GRF (HKU7058/11P) and CRF (HKU8/CRF/11G) of the RGC
of Hong Kong.


\begin{thebibliography}{10}

\bibitem{Berry}
M. V. Berry, Proc. Roy. Soc. London A {\bf 392}, 45 (1984).

\bibitem{Sun} C. P. Sun and
M. L. Ge, Phys. Rev. D {\bf 41}, 1349 (1990); S. L. Zhu and Z. D.
Wang, Phys. Rev. Lett. {\bf 85}, 1076 (2000); S. L. Zhu, Z. D.
Wang, and Y. D. Zhang, Phys. Rev. B {\bf 61}, 1142 (2000).

\bibitem{Wilczek}
F. Wilczek and A. Zee, Phys. Rev. Lett. {\bf 52}, 2111 (1984); A.
Zee, Phys. Rev. A 38, 1 (1988).

\bibitem{GF_review} For a review, see J. Dalibard, F. Gerbier, G. Juzeli\"{u}nas, and P. \"{O}hberg, Rev. Mod.
Phys. {\bf 83}, 1523 (2011).

\bibitem{GF1} R. Dum and M. Olshanii, Phys. Rev. Lett. 76, 1788 (1996).

\bibitem{GF2}
G. Juzeli\={u}nas and P. \"Ohberg, Phys. Rev. Lett. {\bf 93}, 033602
(2004).

\bibitem{GF3}
J. Ruseckas, G. Juzeli\={u}nas, P. \"Ohberg, and M. Fleischhauer,
Phys. Rev. Lett. {\bf 95}, 010404 (2005)

\bibitem{GF4}
G. Juzeli\={u}nas, J. Ruseckas, P. \"Ohberg, and M. Fleischhauer,
Phys. Rev. A {\bf 73}, 025602 (2006).

\bibitem{GF5}
S. L. Zhu, H. Fu, C. J. Wu, S. C. Zhang, and L. M. Duan, Phys. Rev.
Lett. {\bf 97}, 240401 (2006).

\bibitem{GF6} X. J. Liu, X. Liu, L. C. Kwek, and C. H. Oh, Phys. Rev. Lett. {\bf 98}, 026602
(2007).

\bibitem{GF7}T. D. Stanescu, C. Zhang, and V. Galitski,  Phys. Rev. Lett. {\bf 99}, 110403 (2007).

\bibitem{GF8}
K. J. G\"unter, M. Cheneau, T. Yefsah, S. P. Rath, and J. Dalibard,
Phys. Rev. A {\bf 79}, 011604(R) (2009).

\bibitem{GF9}
G. Juzeli\={u}nas, J. Ruseckas, and J. Dalibard, Phys. Rev. A {\bf
81}, 053403 (2010).

\bibitem{GF10} N. R. Cooper, Phys. Rev. Lett. {\bf 106}, 175301
(2011).

\bibitem{GF11} J. D. Sau, R. Sensarma, S. Powell, I. B. Spielman, and S. Das Sarma, Phys. Rev. B {\bf 83}, 140510(R)
(2011).

\bibitem{GF12} D. L. Campbell1, G. Juzeli\={u}nas, and I. B.
Spielman, Phys. Rev. A {\bf 84}, 025602 (2011).

\bibitem{GF13} Z. F. Xu and L. You, arXiv: 1110.5705.

\bibitem{GF14} S. L. Zhu, L.-B. Shao, Z. D. Wang, and L.-M. Duan, Phys. Rev. Lett. {\bf 106}, 100404
(2011); L. Jiang, T. Kitagawa, J. Alicea, A. R. Akhmerov, D. Pekker,
G. Refael, J. I. Cirac, E. Demler, M. D. Lukin, and P. Zoller, {\sl
ibid.} {\bf 106}, 220402 (2011); S. L. Zhu, D. W. Zhang, and Z. D.
Wang, {\sl ibid.} {\bf 102}, 210403 (2009).


\bibitem{NIST_1} Y. J. Lin, R. L. Compton, A. R. Perry, W. D. Phillips, J. V. Porto, and I. B. Spielman, Phys. Rev. Lett. {\bf 102}, 130401
(2009); Y. J. Lin, R. L. Compton, K. Jim\'{e}nez-Garc\'{i}a, J. V.
Porto, and I. B. Spielman, Nature {\bf 462}, 628 (2009); Y. J Lin,
R. L. Compton, K. Jim\'{e}nez-Garc\'{i}a, W. D. Phillips, J. V.
Porto, and I. B. Spielman, Nature Phys., {\bf 7}, 531, (2011).

\bibitem{NIST_SOC} Y.-J. Lin, K. Jim\'{e}nez-Garc\'{i}a, and I. B. Spielman, Nature, {\bf 471}, 83
(2011).

\bibitem{jingzhang} Z. Fu , P. Wang, S. Chai, L. Huang, and J. Zhang, Phys. Rev. A {\bf 84},
043609 (2011).

\bibitem{Bloch2011} M. Aidelsburger, M. Atala, S. Nascimb¨¨ne, S. Trotzky, Y.-A. Chen,
I. Bloch, Phys. Rev. Lett. {\bf 107}, 255301 (2011).

\bibitem{Struck} J. Struck, C. Olschlager, R. Le Targat, P. Soltan-Panahi,
A. Eckardt, M. Lewenstein, P. Windpassinger, and K. Sengstock,
Science {\bf 333}, 996 (2011).

\bibitem{Galitski}
T. D. Stanescu, B. Anderson, and V. Galitski, Phys. Rev. A {\bf 78},
023616 (2008).

\bibitem{Wang}
C. Wang, C. Gao, C. M. Jian, and H. Zhai, Phys. Rev. Lett. {\bf
105}, 160403 (2010).

\bibitem{Jian}
C. M. Jian and H. Zhai, Phys. Rev. B {\bf 84}, 060508 (2011).

\bibitem{Ho}
T. L. Ho and S. Zhang, Phys. Rev. Lett. {\bf 107}, 150403 (2011).

\bibitem{Congjun}
C. J. Wu, I. Mondragon-Shem, and X. F. Zhou, Chin. Phys. Lett. {\bf
28}, 097102 (2011).

\bibitem{Yip}
S. K. Yip, Phys. Rev. A {\bf 83}, 043616 (2011).

\bibitem{Xu}
Z. F. Xu, R. L\"u, and L. You, Phys. Rev. A {\bf 83}, 053602 (2011).

\bibitem{Machida}
T. Kawakami, T. Mizushima, and K. Machida, Phys. Rev. A {\bf 84},
011607 (2011).

\bibitem{Duine}
E. van der Bijl and R.A. Duine, Phys. Rev. Lett. {\bf 107}, 195302
(2011).

\bibitem{Sarang}
S. Gopalakrishnan, A. Lamacraft, and P. M. Goldbart, Phys. Rev. A
{\bf 84}, 061604(R) (2011).

\bibitem{Santos}
S. Sinha, R. Nath, and L. Santos, Phys. Rev. Lett. {\bf 107}, 270401
(2011).

\bibitem{Hu}
H. Hu, B. Ramachandhran, H. Pu, and X. J. Liu, Phys. Rev. Lett. {\bf
108}, 010402 (2012).


\bibitem{Biao}
Q. Zhu, C. Zhang, and B. Wu, arXiv: 1109.5811.

\bibitem{DaSarma}
R. Barnett, S. Powell, T. Grass, M. Lewenstein, and S. Das Sarma,
arXiv: 1109.4945.

\bibitem{SuYi}
Y. Deng, J. Cheng, H. Jing, C.-P. Sun, and S. Yi, arXiv: 1110.0558.

\bibitem{Han}
X.-Q. Xu and J. H. Han, Phys. Rev. Lett. {\bf 107}, 200401 (2011).

\bibitem{Congjun_2}
X. F. Zhou, J. Zhou, and C. J. Wu, Phys. Rev. A {\bf 84}, 063624
(2011) .

\bibitem{Galitski_2}
J. Radi\'{c}, T. Sedrakyan, I. Spielman, and V. Galitski, Phys. Rev.
A {\bf 84}, 063604 (2011).

\bibitem{Merkl}
M. Merkl, A. Jacob, F. E. Zimmer, P. \"Ohberg, and L. Santos, Phys.
Rev. Lett. {\bf 104}, 073603 (2010).

\bibitem{Zhang}
Y. Zhang, L. Mao, and C. Zhang, arXiv: 1102.4045 [Phys. Rev. Lett.
(to be published)].

\bibitem{Zhu}
D. W. Zhang, Z. Y. Xue, H. Yan, Z. D. Wang, and S. L. Zhu, arXiv:
1104.0444 [Phys. Rev. A (to be published)].

\bibitem{Javanaien} J. Javanainen, Phys. Rev. Lett. {\bf 57}, 3164
(1986).

\bibitem{Milburn} G. J. Milburn, J. Corney, E. M. Wright, and D. F.
Walls, Phys. Rev. A {\bf 55}, 4318 (1997).

\bibitem{Smerzi} A. Smerzi, S. Fantoni, S. Giovanazzi, and S. R. Shenoy, Phys. Rev. Lett. {\bf 79}, 4950
(1997); S. Raghavan, A. Smerzi, S. Fantoni, and S. R. Shenoy, Phys.
Rev. A {\bf 59}, 620 (1999).

\bibitem{Giovanazzi} S. Giovanazzi, A. Smerzi, and S. Fantoni, Phys. Rev. Lett. {\bf 84},
4521 (2000).

\bibitem{Levy} S. Levy, E. Lahoud, I. Shomroni, and J. Steinhauer,
Nature {\bf 449}, 579 (2007).

\bibitem{Albiez} M. Albiez, R. Gati, J. F\"{o}lling, S. Hunsmann,
M. Cristiani, and M, K. Oberthaler, Phys. Rev. Lett. {\bf 95},
010402 (2005).

\bibitem{LeBlanc} L. J. LeBlanc, A. B. Bardon, J. McKeever, M. H. T. Extavour, D.
Jervis, J. H. Thywissen, F. Piazza, and A. Smerzi, Phys. Rev. Lett.
{\bf 106}, 025302 (2011).

\bibitem{Inguscio} G. Thalhammer, G. Barontini, L. De Sarlo, J. Catani, F.
Minardi, and M. Inguscio, Phys. Rev. Lett. {\bf 100}, 210402 (2008).

\bibitem{Chang} M.-S. Chang, Q. Qin, W. Zhang, L. You, and M. S.
Chapman, Nat. Phys. {\bf 1}, 111 (2005).

\bibitem{YQLi} X.-Q. Xu, L.-H. Lu, and Y.-Q. Li, Phys. Rev. A {\bf 78}, 043609
(2009).

\bibitem{Pindzola} B. Sun and M. S. Pindzola, Phys. Rev. A {\bf 80}, 033616
(2009).

\bibitem{Pu} H. Pu, W. P. Zhang, and P. Meystre, Phys. Rev. Lett. {\bf 89}, 090401
(2002).

\bibitem{WMLiu} R. Qi, X.-L. Yu, Z. B. Li, and W. M. Liu, Phys. Rev. Lett. {\bf 102}, 185301
(2009).

\bibitem{Chin} C. Chin, R. Grimm, P. Julienne, and E. Tiesinga, Rev. Mod.
Phys. {\bf 82}, 1225 (2010).

\bibitem{YAChen} Y.-A Chen, S. Nascimb\`{e}ne, M. Aidelsburger, M. Atala,
S. Trotzky, I. Bloch,  Phys. Rev. Lett. {\bf 107}, 210405 (2011).


\end{thebibliography}
\end{document}